\documentclass{article}
\usepackage{amsmath}
\usepackage{amssymb}
\usepackage{graphicx}
\usepackage[margin=1.5in]{geometry}

\title{A Crude Analysis of Twitch Plays Pokemon}
\author{Scott Deeann Chen\\
\small{\texttt{chen124@illinois.edu}}}

\date{\vspace{-5ex}}
\begin{document}
\maketitle

\abstract{We model and study the game mechanisms and human behavior of the anarchy mode in Twitch Plays Pokemon with a pure-jump continuous-time Markov process. We computed the winning probability and expected game time for $1$ player and $N$ players and identified when collaboration helps. A numerical plug-in example is also provided.}
\section{Introduction}
Twitch Plays Pokemon \cite{twitchplayspokemon} is a gaming channel that lets viewers collaboratively play a Pokemon Red/Blue game by giving commands through a chatroom, such as "up", "down", "left", "right", "a", "b", "start", or "select". The game has gained a lot of attention in the media \cite{news1}\cite{news2}\cite{news3}, and has accumulated 32 millions views and 184 thousands followers (as of 2014-02-27) since it started on 2014-02-12. There are constantly more than tens of thousands of viewers and players watching and playing the game during this period.

As reported in the game progress document \cite{status}, the game has been progressing much slower than any "rational" single player would achieve, but why would not such a "crowd-sourcing" or "collaboration" scheme be better? How slow would a crowd-sourced game compare to a normal playthrough in terms of the number of players? In this paper, we will discuss the questions above and try to give an estimate for the time growth in terms of the number of players. 

There are two modes in Twitch Plays Pokemon, anarchy and democracy. In anarchy mode, all inputs registered are executed, and in democracy mode, the only the majority voted input is executed. In this report, we will mainly model and discuss the anarchy mode.

Section 2 describes our crude model of a turn-based single player game. Section 3 computes the winning probability and the expected game time under the model for a single player and for $N$ players. Section 4 concludes this paper and discusses future work.

\section{A Crude Model of a Turn-Based Single Player Game}
We simplify a turn-based single player game with the following assumptions:
\begin{enumerate}
\item Each move a player makes is either good or bad.
\item An bad move cancels out a good move, and vise versa.
\item A player wins a game when he/she accumulates $n$ good moves, and a player loses a game when he/she accumulates $m$ bad moves.
\end{enumerate}
We then model the each playthrough as a pure-jump Markov process, as depicted in Figure 1, with $n+m+1$ states, $Z_i$, $i\in\mathbb{Z}$, $-m\leq i\leq n$. State $Z_0$ is the start state, $Z_{n}$ is the winning state, and $Z_m$ is the losing state. Each $Z_i$ represents a game state where $i$ good minus bad moves are made.

\begin{figure}
\includegraphics[width=\textwidth]{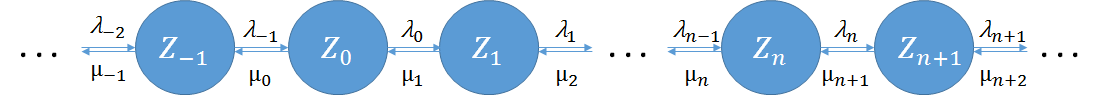}
\caption{The Turn-Based Single Player Game Model}
\end{figure}

In the Markov process, the rate of transition from $Z_i$ to $Z_{i+1}$, making a progress (good move) in game, is $\lambda_i$. The rate of transition from $Z_i$ to $Z_{i-1}$, making a mistake (bad move) in game, is $\mu_{i}=\mu$. For simplicity, we assume $\lambda_i = \lambda$, $\mu_i=\mu$. Then, the corresponding jump process of the Markov process has transition probability $p = \frac{\lambda}{\lambda+\mu}$ from $Z_i$ to $Z_{i+1}$ and $q=1-p= \frac{\mu}{\lambda+\mu}$ from $Z_i$ to $Z_{i-1}$.

\section{Winning Probability and Expected Game Time}
Given the above model, we can then compute the winning probability and expected game time. 

The probability of starting at state $Z_0$ and ending up winning at $Z_n$ is a known result from the gambler's ruin problem with $m$ as the gambler's initial wealth, as the player can make a total of $m$ bad moves before losing. The probability is 
\begin{align}
P &= \frac{1-(\frac{q}{p})^m}{1-(\frac{q}{p})^{n+m}}\\
&=\frac{1-(\frac{\mu}{\lambda})^m}{1-(\frac{\mu}{\lambda})^{n+m}}
\end{align}

The expected game time is the expected number of moves times the expected time spent for each move. The expected number of moves until winning is, again, a known result from the gambler's ruin problem with initial wealth $m$. The expected number of moves is 
\begin{align}
M &= \frac{m}{q-p} - \frac{m+n}{q-p}\times\frac{1-(\frac{q}{p})^m}{1-(\frac{q}{p})^{m+n}}\\
&= (\lambda+\mu)\left[\frac{m}{\mu-\lambda} - \frac{m+n}{\mu-\lambda}\times\frac{1-(\frac{\mu}{\lambda})^m}{1-(\frac{\mu}{\lambda})^{m+n}}\right]\\
\end{align}
Also, according to Markov process properties, the expected time spent at a state $Z_i$ is $t = \frac{1}{\lambda+\mu}$. Therefore, the expected game time is:
\begin{align}
T_g&=Mt\\
&=(\lambda+\mu)\left[\frac{m}{\mu-\lambda} - \frac{m+n}{\mu-\lambda}\times\frac{1-(\frac{\mu}{\lambda})^m}{1-(\frac{\mu}{\lambda})^{m+n}}\right]\frac{1}{\lambda+\mu}\\
&=\frac{m}{\mu-\lambda} - \frac{m+n}{\mu-\lambda}\times\frac{1-(\frac{\mu}{\lambda})^m}{1-(\frac{\mu}{\lambda})^{m+n}}
\end{align}

The expected game time is for a single playthrough, but not for a winning playthrough, as there is a probability that the player will lose. To compute the expected game time to win the game, we view each playthrough as a geometric trial with success probability $P$. The expected game time to win the game is then the expected number of playthroughs times the expected time for each playthrough.

\begin{align}
T_w &= \frac{1}{P}\times T_g\\
&=\frac{1-(\frac{\mu}{\lambda})^{n+m}}{1-(\frac{\mu}{\lambda})^m}\left[\frac{m}{\mu-\lambda} - \frac{m+n}{\mu-\lambda}\times\frac{1-(\frac{\mu}{\lambda})^m}{1-(\frac{\mu}{\lambda})^{m+n}}\right]\\
&=\frac{m}{\mu-\lambda}\times\frac{1-(\frac{\mu}{\lambda})^{m+n}}{1-(\frac{\mu}{\lambda})^{m}} - \frac{m+n}{\mu-\lambda}
\end{align}

To compute $P$ and $T_w$ for the single player and $N$ player case, we will characterize $\lambda$ and $\mu$ by modeling game mechanics and human behavior as below:
\begin{enumerate}
\item All inputs from all users are registered and executed immediately but observed after $T_d$ seconds.
\item A human needs $T_h$ to react to a game state change, including understanding the change, planing for moves, etc. The time $T_h$ is a random variable and follows an exponential distribution with mean $\lambda_h^{-1}$. 
\item A human only reacts once to the current observed state of the game.
\item Duplicate inputs make a correct input incorrect.
\end{enumerate}
Also, we define the quality $q$ of a player to be the probability of the player's input to be correct. 

We are then ready to consider two scenarios.
\begin{enumerate}
\item A single player.
\item A total of $N$ players.
\end{enumerate}

\subsection{A Single Player}
For a single player, the rate of getting a correct input is $\lambda = q\lambda_h$, and the rate of getting a incorrect input is $\mu = (1-q)\lambda_h$.

The winning probability of the player is:

\begin{align}
P_1 &= 
\left\{ \begin{array}{ll}
 (\frac{p}{1-p})^n & \mbox{ if $ p \leq 1-p$} \\
1 &\mbox{ otherwise}
       \end{array} \right.\\
&=\left\{ \begin{array}{ll}
 (\frac{q\lambda_h}{1-q\lambda_h})^n & \mbox{ if $ q\lambda_h \leq 1-q\lambda_h$} \\
1 &\mbox{ otherwise}
       \end{array} \right.
\end{align}
The winning probability depends on the quality of the user. The higher the quality is, the higher the winning probability is. 

When we consider the expected total game time, we have to take $T_d$ the observation delay into account. Luckily, this is not a bid deal. The expected total game time is:
\begin{align}
T_{g1}&= M(t+T_d)\\
&=\frac{n(\lambda+\mu)}{\lambda-\mu}(\frac{1}{\lambda+\mu}+T_d)\\
&=\frac{n}{\lambda-\mu} + \frac{n(\lambda+\mu)}{\lambda-\mu}T_d\\
&=\frac{n}{\lambda_h(2q-1)}(1+\lambda_hT_d).
\end{align}
No surprise here. The total game time is linear to the length of the winning sequence $n$, the average reaction time of a human $\frac{1}{\lambda_h}$, and the observation delay $T_d$. 

In the case where the rate $\lambda_h$ is high, i.e., the user thinks, plans, and inputs fast, we have 
\begin{align}
\lim\limits_{\lambda_h\to\infty} = \frac{nT_d}{2q-1}.
\end{align}
The game time $T_{g1}$ is dominated by the delay time and the length of the winning sequence $n$. In the other case where $\lambda_h$ is small, we have $ \lambda_hT_d \ll 1$ and \begin{align}
T_g\approx \frac{n}{\lambda_h(2q-1)}.
\end{align}
The game time $T_g$ is dominated by the length of the winning sequence $n$ and the input rate $\lambda_h$.
\subsection{A Total of $N$ Players}
Now comes the tricky part, when $N$ players are playing simultaneously without collaboration, they can easily make duplicated inputs that reverse the game progress (by assumption). A total of $N$ players make a progress only when the first input is correct and nothing is input within the next $T_d$ seconds. 

Without loss of generality, denote $T_{hi}$ the time the $i$th player takes to make an input, and $T_{hi}<T_{hj}$ when $i<j$, $i\neq j$, and $1\leq i,j \leq N$. The exponential distribution rate between $T_{hi}$ and $T_{h(i+1)}$ is $\lambda_h (N-i)$, $1\leq i \leq N-1$, as there are $(N-i)$ players trying to input during the interval.
By the memoryless property of exponential distributions, the probability of the event is then:
\begin{align}
P_c&=P[T_{h2}>T_{h1}+T_d|T_{h2}>T_{h1}]\\
&=P[T_{h2}>T_d]\\
&= e^{-\lambda_h(N-1)T_d}
\end{align}

We can then compute the rates $\lambda$ and $\mu$. The difference between the $N$ player case and the one player case is that there is a chance  that a correct input will be incorrect due to a duplicate input, and could be addressed by adjusting the quality of the input $q'=q*P_c$. Also, the rates between time $0$ and $T_{h1}$ is $\lambda = qN\lambda_m$ and $\mu=(1-q)N\lambda_m$, since there are $N$ players with quality $q$ inputting at rate $\lambda_m$. 

Therefore, the winning probability is
\begin{align}
P_1 &= 
\left\{ \begin{array}{ll}
 (\frac{p}{1-p})^n & \mbox{ if $ p \leq 1-p$} \\
1 &\mbox{ otherwise}
       \end{array} \right.\\
       &=\left\{ \begin{array}{ll}
        (\frac{qN\lambda_he^{-\lambda_h(N-1)T_d}}{1-qN\lambda_he^{-\lambda_h(N-1)T_d}})^n & \mbox{ if $qN\lambda_he^{-\lambda_h(N-1)T_d} \leq 1-qN\lambda_he^{-\lambda_h(N-1)T_d}$} \\
       1 &\mbox{ otherwise}
              \end{array} \right.
\end{align}
Note the increase in $N$ decreases $qN\lambda_he^{-\lambda_h(N-1)T_d}$ and makes a guaranteed win non-guaranteed.

We can also calculate the he expected total game time as below:
\begin{align}
T_{gN}&= M(t+T_d)\\
&=\frac{n(\lambda+\mu)}{\lambda-\mu}(\frac{1}{\lambda+\mu}+T_d)\\
&=\frac{n}{\lambda-\mu} + \frac{n(\lambda+\mu)}{\lambda-\mu}T_d\\
&=\frac{n}{N\lambda_h(2qe^{-\lambda_h(N-1)T_d}-1)}(1+N\lambda_hT_d).
\end{align}
The expected game time grows exponentially with $N$ in the region where $2qe^{-\lambda_h(N-1)T_d} > 1$ and becomes infinite when $2qe^{-\lambda_h(N-1)T_d} \leq 1$. By plugging in some numbers: 
\begin{itemize}
\item $q=0.99$: everybody is an expert,
\item $T_d=0.15$: human vision reaction time,
\item $\lambda_h= 10^{-1}$: 10 seconds to evaluate the situation and type up an input.
\item $n=100$: The length of the winning sequence is $100$.
\end{itemize}
The maximum $N$ such that $2qe^{-\lambda_h(N-1)T_d} > 1$ holds is $N=46$. This suggest that if $46$ experts play the game simultaneously, while they are guaranteed to win, the expected game time will be very long, and more than $46$ players will not guarantee a win.

Figure 2 shows the expected game time vs. the number of players for the above numerical example. We make the following observations.
\begin{figure}[h]
\includegraphics[width=\textwidth]{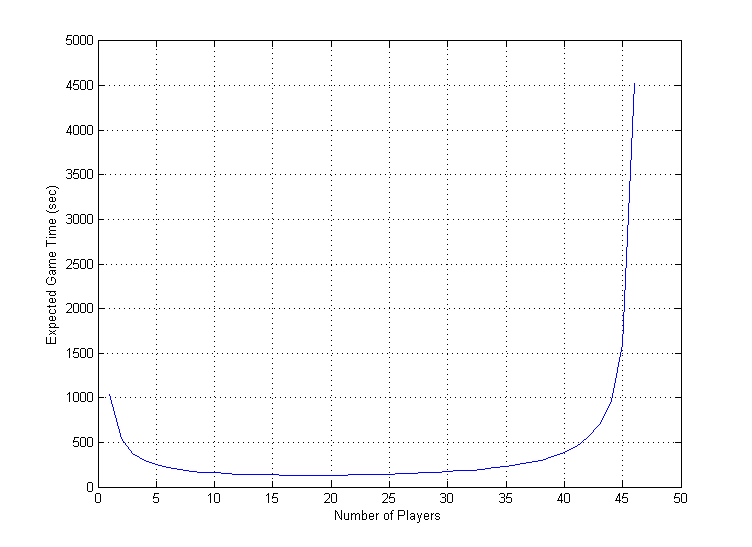}
\caption{A numerical example of expected game time vs. number of players for $q=0.99$, $T_d=0.15$, $\lambda_h= 10^{-1}$, $n=100$.}
\end{figure}
Playing together, even without collaboration, does help! The expected game time drops around $7$ folds from one player to $19$ players. The intuition is that for difficult problems, even if everybody worked independently, others are able to learn from the first person who solved the problem. Collaboration helps especially in problems where $\lambda_h$ and $T_d$ is small. This setting means that the problem is hard to solve, yet once it is solved, others can learn the solution easily and then proceed, just like the scientific research community.

When the number of players are large, everybody repeats the same effort and destroys what has already been done. This can be observed in Twitch Plays Pokemon when people try to navigate through a maze through anarchy mode.

\section{Discussion and Future Work}
In this report, we modeled the game mechanics and human behavior of Twitch Plays Pokemon anarchy mode by a pure-jump continuous-time Markov process. Numerical plug-in results are presented and showed two findings: 1) crowd sourcing without collaboration does help and 2) anarchy mode of Twitch Plays Pokemon is not a good idea to reduce game play time (it is fun to watch, though).

Our future work will focus on two directions. First, we want to remove unrealistic assumptions, such as 1) "no trolls!" The model is not complete without Internet trolls, and 2) There could be a third type of move in addition to correct and incorrect ones. Second, we want to investigate how much improve in expected game time will democracy mode have and how will the expected game time grow or shrink with $N$ in democracy mode.


\newpage
\bibliographystyle{IEEEtran}
\bibliography{ref}

\begin{thebibliography}{1}
\providecommand{\url}[1]{#1}
\csname url@samestyle\endcsname
\providecommand{\newblock}{\relax}
\providecommand{\bibinfo}[2]{#2}
\providecommand{\BIBentrySTDinterwordspacing}{\spaceskip=0pt\relax}
\providecommand{\BIBentryALTinterwordstretchfactor}{4}
\providecommand{\BIBentryALTinterwordspacing}{\spaceskip=\fontdimen2\font plus
\BIBentryALTinterwordstretchfactor\fontdimen3\font minus
  \fontdimen4\font\relax}
\providecommand{\BIBforeignlanguage}[2]{{%
\expandafter\ifx\csname l@#1\endcsname\relax
\typeout{** WARNING: IEEEtran.bst: No hyphenation pattern has been}%
\typeout{** loaded for the language `#1'. Using the pattern for}%
\typeout{** the default language instead.}%
\else
\language=\csname l@#1\endcsname
\fi
#2}}
\providecommand{\BIBdecl}{\relax}
\BIBdecl

\bibitem{twitchplayspokemon}
\BIBentryALTinterwordspacing
``Twitch {P}lays {P}okemon,'' {A}ccessed: 2014-03-13. [Online]. Available:
  \url{http://www.twitch.tv/twitchplayspokemon}
\BIBentrySTDinterwordspacing

\bibitem{news1}
\BIBentryALTinterwordspacing
H.~Tsukayama, ``How 120,000 players managed to play one epic game of
  {P}okemon,'' {A}ccessed: 2014-03-13. [Online]. Available:
  \url{http://www.washingtonpost.com/blogs/the-switch/wp/2014/02/21/how-120000-players-managed-to-play-one-epic-game-of-pokemon/}
\BIBentrySTDinterwordspacing

\bibitem{news2}
\BIBentryALTinterwordspacing
M.~R. Dickey, ``Here's what happens when 100,000 people try to play a game of
  {P}okémon at the same time,'' {A}ccessed: 2014-03-13. [Online]. Available:
  \url{http://www.businessinsider.com/twitchs-servers-overloaded-pokemon-2014-2}
\BIBentrySTDinterwordspacing

\bibitem{news3}
\BIBentryALTinterwordspacing
R.~Rigney, ``Millions unite for a game of {P}okémon. you can help (or
  troll),'' {A}ccessed: 2014-03-13. [Online]. Available:
  \url{http://www.wired.com/2014/02/twitch-plays-pokemon/}
\BIBentrySTDinterwordspacing

\bibitem{status}
\BIBentryALTinterwordspacing
``Twitch plays {P}okemon google document main page,'' accessed: 2014-03-13.
  [Online]. Available:
  \url{https://sites.google.com/site/twitchplayspokemonstatus/}
\BIBentrySTDinterwordspacing

\end{thebibliography}


\end{document}